\documentclass[sigconf]{acmart}

\usepackage{todonotes}

\newcommand{\gr}[1]{\textcolor{gray}{#1}}

\copyrightyear{2024} 
\acmYear{2024} 
\setcopyright{rightsretained} 
\acmConference[SIGIR '24]{Proceedings of the 47th International ACM SIGIR Conference on Research and Development in Information Retrieval}{July 14--18, 2024}{Washington, DC, USA}
\acmBooktitle{Proceedings of the 47th International ACM SIGIR Conference on Research and Development in Information Retrieval (SIGIR '24), July 14--18, 2024, Washington, DC, USA}
\acmDOI{10.1145/3626772.3657856}
\acmISBN{979-8-4007-0431-4/24/07}

\begin{document}

\title{A Reproducibility Study of PLAID}

\author{Sean MacAvaney}
\orcid{0000-0002-8914-2659}
\affiliation{%
  \institution{University of Glasgow}
  \city{Glasgow}
  \country{United Kingdom}
}
\email{sean.macavaney@glasgow.ac.uk}

\author{Nicola Tonellotto}
\orcid{0000-0002-7427-1001}
\affiliation{%
  \institution{University of Pisa}
  \city{Pisa}
  \country{Italy}
}
\email{nicola.tonellotto@unipi.it}

\renewcommand{\shortauthors}{MacAvaney and Tonellotto}

\begin{abstract}
The PLAID (Performance-optimized Late Interaction Driver) algorithm for ColBERTv2 uses clustered term representations to retrieve and progressively prune documents for final (exact) document scoring. In this paper, we reproduce and fill in missing gaps from the original work. By studying the parameters PLAID introduces, we find that its Pareto frontier is formed of a careful balance among its three parameters; deviations beyond the suggested settings can substantially increase latency without necessarily improving its effectiveness. We then compare PLAID with an important baseline missing from the paper: re-ranking a lexical system. We find that applying ColBERTv2 as a re-ranker atop an initial pool of BM25 results provides better efficiency-effectiveness trade-offs in low-latency settings. However, re-ranking cannot reach peak effectiveness at higher latency settings due to limitations in recall of lexical matching and provides a poor approximation of an exhaustive ColBERTv2 search. We find that recently proposed modifications to re-ranking that pull in the neighbors of top-scoring documents overcome this limitation, providing a Pareto frontier across all operational points for ColBERTv2 when evaluated using a well-annotated dataset. Curious about why re-ranking methods are highly competitive with PLAID, we analyze the token representation clusters PLAID uses for retrieval and find that most clusters are predominantly aligned with a single token and vice versa. Given the competitive trade-offs that re-ranking baselines exhibit, this work highlights the importance of carefully selecting pertinent baselines when evaluating the efficiency of retrieval engines.

\vspace{0.6em}
\hspace{1.8em}\includegraphics[width=1.25em,height=1.25em]{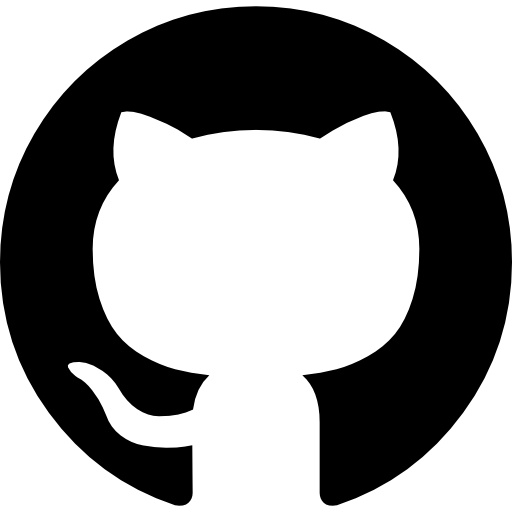}\hspace{.3em}
\parbox[c]{\columnwidth}
{
    \vspace{-.55em}
    \href{https://github.com/seanmacavaney/plaidrepro}{\nolinkurl{https://github.com/seanmacavaney/plaidrepro}}
}
\vspace{-1.2em}
\end{abstract}




\maketitle

\section{Introduction}

Relevance ranking is a central task in information retrieval. Numerous classes of models exist for the task, including lexical~\cite{DBLP:conf/trec/RobertsonWJHG94}, dense~\cite{DBLP:conf/emnlp/KarpukhinOMLWEC20}, learned sparse~\cite{DBLP:conf/ecir/NguyenMY23}, and late interaction~\cite{DBLP:conf/sigir/KhattabZ20}. While efficient exact top-$k$ retrieval algorithms exist for lexical and learned sparse retrieval systems (e.g., BlockMaxWAND~\cite{DBLP:conf/sigir/DingS11}), dense and late interaction methods either need to perform expensive exhaustive scoring over the entire collection or resort to an approximation of top-$k$ retrieval. A myriad of approximate $k$-nearest-neighbor approaches are available for (single-representation) dense models (e.g., HNSW~\cite{DBLP:journals/pami/MalkovY20}). However, these approaches generally do not apply directly to late interaction scoring mechanisms, so bespoke retrieval algorithms for late interaction models have been proposed.

\begin{figure}
\centering
\includegraphics[scale=0.65]{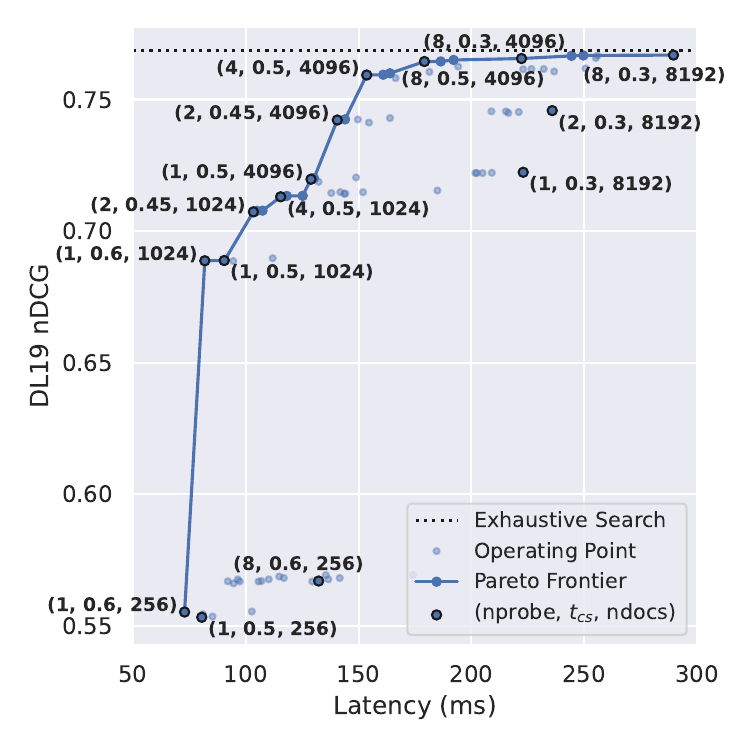}
\vspace{-2em}
\caption{The Pareto frontier of PLAID for ColBERTv2 on TREC DL 2019 over the three parameters it introduces (\texttt{nprobe}, $t_{cs}$, and \texttt{ndocs}). Several operational points are labeled to highlight the interdependence of PLAID's parameters.}
\label{fig:plaid-pareto}
\end{figure}

PLAID (Performance-optimized Late Interaction Driver)~\cite{plaid} is one such retrieval algorithm. It is designed to efficiently retrieve and score documents for ColBERTv2~\cite{colbertv2}, a prominent late interaction model. PLAID first performs coarse-grained retrieval by matching the closest ColBERTv2 centroids (used for compressing term embeddings) to the query term embeddings. It then progressively filters the candidate documents by performing finer-grained estimations of a document's final relevance score. These filtering steps are controlled by three new parameters, which are discussed in more detail in Section~\ref{sec:background}.

The original PLAID paper answered several important research questions related to the overall effectiveness and efficiency compared to ColBERTv2's default retriever, the effect of the individual filtering stages, and its transferability. However, it also left several essential questions unanswered. This reproducibility paper aims to reproduce the core results of the paper and answer several additional questions. First, we explore the effect of PLAID's new parameters to better understand the practical decisions one must make when deploying a PLAID engine. Then we explore and report an important missing baseline (re-ranking a lexical retrieval system), which has been shown to be a highly-competitive approach for dense systems~\cite{DBLP:journals/corr/abs-2110-06051,DBLP:conf/sigir/KulkarniMGF23}. Throughout our exploration, we also answer questions about how well PLAID applies to a dataset with many known relevant documents and how well it approximates an exhaustive ColBERTv2 search.

We find that PLAID's parameters need to be carefully set in conjunction with one another to avoid sub-optimal tradeoffs between effectiveness and efficiency. As shown in Figure~\ref{fig:plaid-pareto}, PLAID's Pareto frontier is a patchwork of parameter settings; changing one parameter without corresponding changes to the others can result in slower retrieval with no change to effectiveness. Further, we find that re-ranking lexical search results provides better efficiency-effectiveness trade-offs than PLAID in low-latency settings. For instance, competitive results can be achieved in a single-threaded setup in as low as 7ms/query with re-ranking, compared to 73ms/query for PLAID. Through an analysis of the token clusters, we confirm that a large proportion of tokens predominantly perform lexical matching, explaining why a lexical first stage is so competitive. We feel that our study provides important operational recommendations for those looking to use ColBERTv2 or similar models, both with and without the PLAID algorithm.

\section{Background and Preliminaries}\label{sec:background}

The late interaction class of ranking models applies a lightweight query-token to document-token ``interaction'' operator atop a contextualized text encoder to estimate relevance between a query and a document. Perhaps most well-known is the ColBERT model~\cite{DBLP:conf/sigir/KhattabZ20}, which applies a maximum-similarity operator over a pretrained transformer-based language model---though other late interaction operators (e.g., \cite{DBLP:conf/sigir/MacAvaneyYCG19,DBLP:conf/emnlp/ZhouD21}) and contextualization strategies (e.g.,~\cite{DBLP:conf/wsdm/DaiXC018,DBLP:conf/ecai/HofstatterZH20}) have also been proposed. Due to the nature of their scoring mechanism, late interaction models cannot efficiently identify the exact top-$k$ search results without an exhaustive scan over all documents,\footnote{In contrast, learned sparse models can be stored in inverted indices and retrieved using algorithms such as BlockMax-WAND~\cite{DBLP:conf/sigir/DingS11}.} nor can they directly use established approximate nearest neighbor algorithms.\footnote{In contrast, single-representation dense retrieval models can be indexed and retrieved using algorithms such as HNSW~\cite{DBLP:journals/pami/MalkovY20}.} Early work in late interaction approaches (e.g.,~\cite{DBLP:conf/sigir/MacAvaneyYCG19,DBLP:conf/ecai/HofstatterZH20,DBLP:conf/wsdm/DaiXC018}) overcame this limitation through a re-ranking strategy, wherein a candidate set of documents are identified using an efficient first-stage lexical retriever such as BM25~\cite{DBLP:conf/trec/RobertsonWJHG94}. \citet{DBLP:conf/sigir/KhattabZ20} identified that this re-ranking strategy may be sub-optimal, since the (lexically-matched) first stage results may not be aligned with those that the model will rank highly. Therefore, they proposed using approximate $k$-nearest neighbour search over the token representations to identify documents to score instead.

To deal with the considerable space requirements to store the pre-computed representations of document tokens, ColBERTv2~\cite{colbertv2} implements a clustering solution to identify document token centroids that can be used to decompose a document token representation as a sum of a centroid and a quantized residual vector, reducing the storage requirements by one order of magnitude w.r.t. the original ColBERT. These cluster centroids can serve as proxies of document tokens~\cite{plaid,colbertprf1,colbertprf2}.

\begin{figure}
\centering
\includegraphics[scale=0.36]{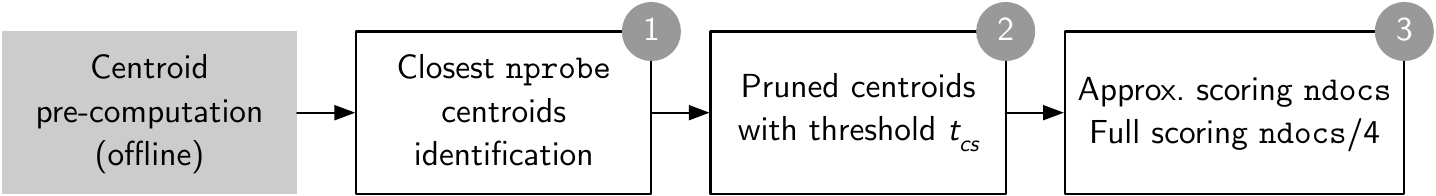}
\vspace{-2em}
\caption{The logical phases composing the candidate documents identification procedure in PLAID.}
\label{fig:phases}
\end{figure}

PLAID~\cite{plaid} further builds upon the centroids of ColBERTv2 to improve retrieval efficiency. PLAID selects and then progressively filters out candidate documents through three distinct phases, as illustrated in Figure~\ref{fig:phases}. Firstly, given an encoded query token representation, its closest document token centroids are computed. The corresponding document identifiers are retrieved and merged together into a candidate set. The number of closest centroids to match per query token is a hyperparameter called \texttt{nprobe}. Naturally, the initial pool of documents increases in size as \texttt{nprobe} increases. Secondly, the set of candidate centroids is pruned by removing all centroids whose maximum similarity w.r.t. all query tokens is smaller than a threshold parameter $t_{cs}$. Next, the pruned set of centroids is further pruned by selecting the top \texttt{ndocs} documents based on relevance scores computed with the late interactions mechanism on the unpruned centroids. Then, the top \texttt{ndocs}/4 approximately-scored documents are fully scored by decompressing the token representations and computing the exact ColBERTv2 relevance scores.
Note that PLAID introduces a total of three hyperparameters, namely \texttt{nprobe}, $t_{cs}$, and \texttt{ndocs}. Although three suggested configurations of these settings were provided by the original PLAID paper, it does not explore the effects and inter-dependencies between them.

\section{Core Result Reproduction}

We begin by reproducing the core results of PLAID. Specifically, we test that retrieving using PLAID's recommended operational points provides the absolute effectiveness and relative efficiency presented in the original paper. Given the limitation that the original paper experimented with sparsely-labeled evaluation sets, we test one sparsely-labeled dataset from the original paper and one dataset with more complete relevance assessments. We also add a new measurement that wasn't explored in the original work---the Rank-Biased Overlap (RBO)~\cite{DBLP:journals/tois/WebberMZ10} with an exhaustive ColBERTv2 search---to test how good of an approximation PLAID is with respect to a complete search.

Our experimental setup, detailed in the following section, includes both elements of both reproducibility and replicability per ACM's definitions,\footnote{\url{https://www.acm.org/publications/policies/artifact-review-and-badging-current}} since we are a different team using some of the same artifacts (code, model, datasets, etc.), while also introducing other changes to the experimental setup (added an evaluation dataset, new measures, etc.).

\subsection{Experimental Setup}\label{sec:repro_exp}

\textbf{Model and Code.}
We reproduce PLAID starting form the released ColBERTv2 checkpoint\footnote{\url{https://downloads.cs.stanford.edu/nlp/data/colbert/colbertv2/colbertv2.0.tar.gz}} and the PLAID authors' released codebase.\footnote{\url{https://github.com/stanford-futuredata/ColBERT}} We release our modified version of the code and scripts to run our new experiments.

\begin{table}
\centering
\caption{Suggested operational points from the original PLAID paper. We refer to them as (a), (b), and (c).}
\vspace{-1em}
\label{tab:repro_settings}
\begin{tabular}{lrrr}
\toprule
System & \texttt{nprobe} & $t_{cs}$ & \texttt{ndocs} \\
\midrule
PLAID (a) & 1 & 0.50 &  256 \\
PLAID (b) & 2 & 0.45 & 1024 \\
PLAID (c) & 4 & 0.40 & 4096 \\
\bottomrule
\end{tabular}
\vspace{-1em}
\end{table}

\textbf{Parameters.}
We use PLAID's recommended settings for the \texttt{nprobe}, $t_{cs}$, and \texttt{ndocs} parameters, as shown in Table~\ref{tab:repro_settings}. We refer to these operational settings as (a), (b), and (c) for simplicity, where each setting progressively filters \textit{fewer} documents. PLAID performs a final top $k$ selection at the end of the process (i.e., after fully scoring and sorting the filtered documents). We recognize that this step is unnecessary and only limits the \textit{apparent} result set size. Therefore, in line with typical IR experimental procedures, we wet $k=1000$ across all settings. We also use the suggested settings of \texttt{nbits}=2 and \texttt{nclusters}=$2^{18}$.

\textbf{Baselines.} We compare directly against the results reported by the original PLAID paper for our experimental settings (Table 3 in their paper). We further conducted an exhaustive search over ColBERTv2\footnote{I.e., fully scoring all documents. We modified the codebase to support this option.} to better contextualize the results and support the measurement of rank-biased overlap (described below).

\textbf{Datasets.} We evaluate on the MS MARCO v1 passage development dataset~\cite{DBLP:conf/nips/NguyenRSGTMD16,DBLP:conf/sigir/CraswellMYCL21}, which consists of 6,980 queries with sparse relevance assessments (1.1 per query). To make up for the limitations of these assessments, we also evaluate using the more comprehensive TREC DL 2019 dataset~\cite{DBLP:journals/corr/abs-2003-07820}, which consists of 43 queries with 215 assessments per query. In line with the official task guidelines and the original PLAID paper, we do not augment the MS MARCO passage collection with titles~\cite{DBLP:conf/sigir/LassanceC23}.

\textbf{Measures.} For MS MARCO Dev, we evaluate using the official evaluation measure of mean Reciprocal Rank at depth 10 (RR@10), using MS MARCO's provided evaluation script. To understand the overall system's ability to retire the relevant passage, we measure the recall at depth 1000 (R@1k), which is also frequently used for the evaluation of Dev. To test how well PLAID approximates an exhaustive search, we measure Rank Biased Overlap (RBO)~\cite{DBLP:journals/tois/WebberMZ10}, with a persistence of 0.99. We measure efficiency via the mean response time using a single CPU thread over the Dev set in milliseconds per query (ms/q). In line with the original paper, we only measure the time for retrieval, ignoring the time it takes to encode the query (which is identical across all approaches). For TREC DL 2019, we evaluate the official measure of nDCG@10, alongside nDCG@1k to test the quality of deeper rankings and R@1k to test the ability of the algorithm to identify all known relevant passages to a given topic. Following standard conventions on TREC DL 2019, we use a minimum relevance score of 2 when computing recall. We use pytrec\_eval~\cite{DBLP:conf/sigir/GyselR18} to compute these measurements.

\textbf{Hardware.}
The original PLAID paper evaluated multiple hardware configurations, including single-CPU, multi-CPU, and GPU settings. Given the algorithm's focus on efficiency, we exclusively use a single-threaded setting, recognizing that most parts of the algorithm can be trivially parallelized on either CPU or GPU. Also as was done in the original work, we load all embeddings into memory, eliminating the overheads of reading from disk. We conducted our experiments using a machine equipped with a 3.4GHz AMD Ryzen 9 5950X processor. (The original paper used a 2.6 GHz Intel Xeon Gold 6132 processor.)

\subsection{Results}

\begin{table}
\centering
\caption{Core reproduction results. \gr{NR}: Value was Not Reported in the original paper. \gr{N/A}: The latency for an exhaustive search over ColBERTv2 is excessive and not applicable to this study.}
\vspace{-0.5em}
\label{tab:repro_res}
\scalebox{0.87}{
\begin{tabular}{crrrrrrr}
\toprule
&\multicolumn{4}{c}{MS MARCO Dev} & \multicolumn{3}{c}{TREC DL 2019} \\
\cmidrule(lr){2-5}
\cmidrule(lr){6-8}
 & RR@10 & R@1k & RBO & ms/q & nDCG@10 & nDCG@1k & R@1k \\
\midrule
\multicolumn{8}{c}{\bf PLAID Reproduction} \\
(a) & 0.394 & 0.833 & 0.612 & 80.5 & 0.739 & 0.553 & 0.555 \\
(b) & 0.397 & 0.933 & 0.890 & 103.4 & 0.745 & 0.707 & 0.786 \\
(c) & 0.397 & 0.975 & 0.983 & 163.9 & 0.745 & 0.760 & 0.871 \\
\midrule
\multicolumn{8}{c}{\bf Original PLAID Results} \\
(a) & 0.394 & \gr{NR} & \gr{NR} & 185.5 & \gr{NR} & \gr{NR} & \gr{NR} \\
(b) & 0.398 & \gr{NR} & \gr{NR} & 222.3 & \gr{NR} & \gr{NR} & \gr{NR} \\
(c) & 0.398 &   0.975 & \gr{NR} & 352.3 & \gr{NR} & \gr{NR} & \gr{NR} \\
\midrule
\multicolumn{8}{c}{\bf Exhaustive ColBERTv2 Search} \\
- & 0.397 & 0.984 & 1.000 & \gr{N/A} & 0.745 & 0.769 & 0.894 \\
\bottomrule
\end{tabular}
}
\end{table}

Table~\ref{tab:repro_res} presents the results of our core reproduction study. We start by considering the effectiveness reported on MS MARCO Dev. We see virtually no difference across all three operational points in terms of the precision-oriented RR@10 measure.\footnote{We note that tools such as \texttt{repro\_eval}~\cite{DBLP:conf/sigir/Breuer0FMSSS20} are available to provide more fine-grained comparisons of individual rankings. In our study, we are primarily interested in the overall effectiveness, rather than the precise ordering of the results used to achieve these scores.} In terms of efficiency, our absolute latency measurements are lower, though this is not surprising given that we are using a faster CPU. The approximate relative differences between each of the operational points are similar, however, e.g., operational point (b) provides a 37\% speedup over (c) in both the original paper and our reproduction. Regarding R@1k and RBO, we see similar trends to that of RR@10: as the operational settings collectively consider more documents for final scoring, the measures improve. These results demonstrate that PLAID is working as expected: when more documents are considered, PLAID identifies a larger number of relevant documents (R@1k increases) and also produces a better approximation of an exhaustive ColBERTv2 search (RBO increases).

When considering the results on TREC DL 2019, we observe similar trends to the Dev results. The precision-focused nDCG@10 measure improves slightly from (a) to (b), while nDCG@1k and R@1k exhibit larger improvements across the settings due to the improved recall of the system. These results help further demonstrate PLAID's robustness in different evaluation settings.

In summary, we are able to reproduce PLAID's core results (in terms of precision and efficiency) successfully on a single CPU setting. We further validate that the trends hold when measuring PLAID with recall-oriented measures and when evaluating PLAID on a dataset with more complete relevance assessments. However, there are still several limitations with the original evaluation. Although we know three settings in which PLAID's parameters can work together to deliver efficient retrieval, we do not understand the effect of each individually. Further, although PLAID retrieval is quite fast in an absolute sense (down to around 80ms/query on a single CPU core), we do not know how well this compares to highly-competitive re-ranking systems. These limitations are addressed in the following sections.

\section{Parameter Study}\label{sec:params}

\begin{figure*}
\centering
\includegraphics[scale=0.78]{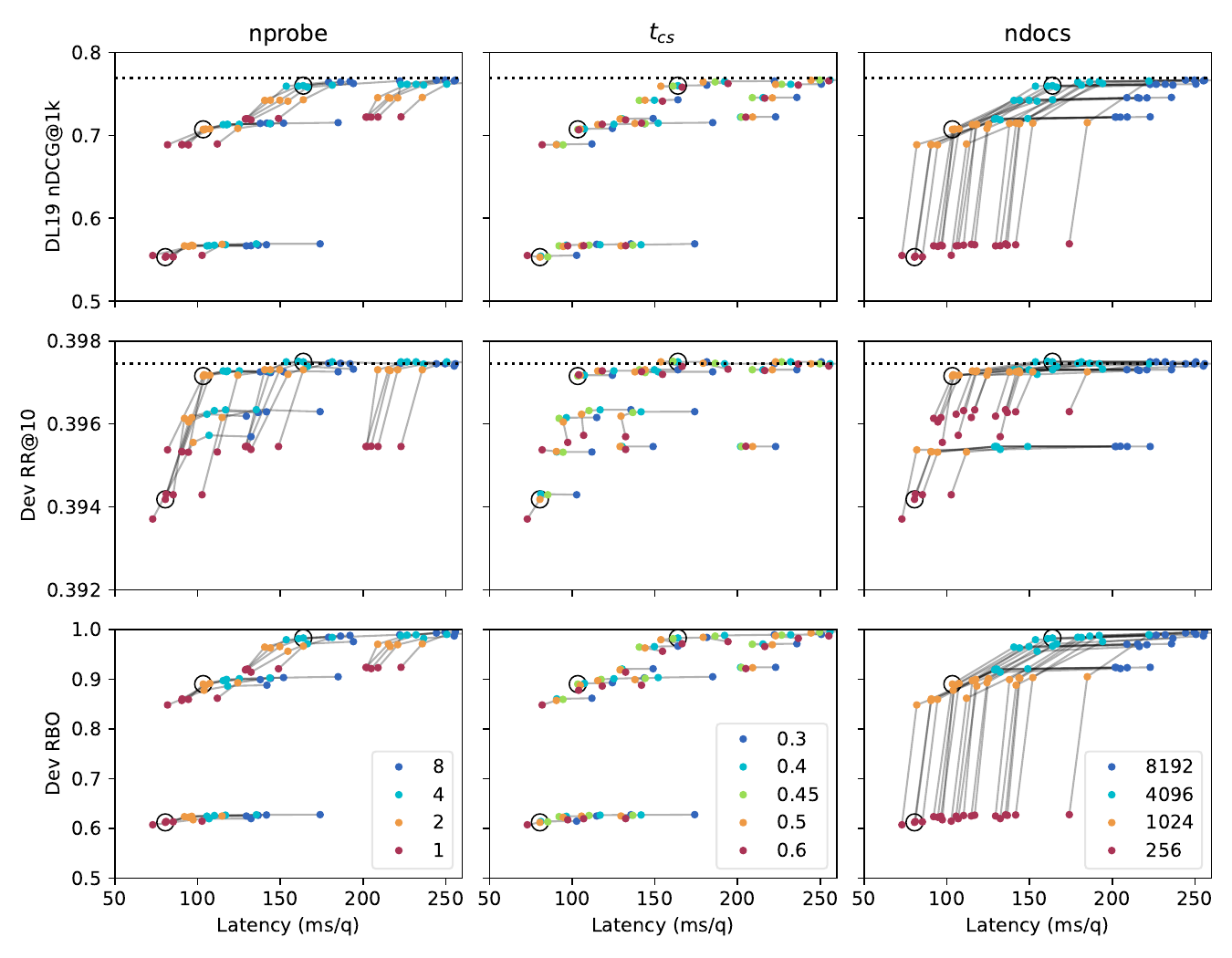}
\vspace{-2em}
\caption{Results of our study of PLAID's parameters \texttt{nprobe}, $t_{cs}$, and \texttt{ndocs}. Each row plots the same data points, with the colors representing each parameter value and the lines between them showing the effect with the other two parameters held constant. The dotted line shows the results of an exhaustive search, and the circled points highlight the three recommended settings from the original paper.}
\label{fig:plaid-params}
\end{figure*}

Recall that PLAID introduces three new parameters: \texttt{nprobe} (the number of clusters retrieved for each token), $t_{cs}$ (the centroid pruning threshold), and \texttt{ndocs} (the maximum number of documents returned after centroid interaction pruning). Although the original paper suggested three settings for these parameters (see Table~\ref{tab:repro_settings}), it did not explain how these parameters were selected or how each parameter ultimately affects retrieval effectiveness or efficiency. In this section, we fill this gap.

\subsection{Experimental Setup}

We extend the experimental setup from our core reproduction study presented in Section~\ref{sec:repro_exp}. We then performed a grid search over the following parameter settings: $\texttt{nprobe}\in\{1, 2, 4, 8\}$, $t_{cs}\in\{0.3, 0.4, 0.45, 0.5, 0.6\}$, and $\texttt{ndocs}\in\{256, 1024, 4096, 8192\}$.

\looseness -1 This set of parameters was initially seeded by performing a grid search over the suggested parameter settings. Given that \texttt{nprobe} already includes the minimum value of 1, we extended it to 8 to check if introducing even more candidate documents from the first stage helps. For $t_{cs}$, we extended the parameter search in both directions: down to 0.3 (filtering out fewer documents based on the centroid scores) and up to 0.6 (filtering out more documents). Finally, we extended \texttt{ndocs} up to 8192, based on our observations that low values of this parameter (e.g., 256) substantially harm effectiveness.

We also asked the PLAID authors about anything else to tweak with PLAID to maximize effectiveness or efficiency. They told us that these three parameters have the most substantial effect. Meanwhile, the indexing-time parameters of \texttt{nbits} and \texttt{nclusters} can also affect the final performance. However, see these two indexing parameters as settings of the ColBERTv2 \textit{model} rather than the PLAID \textit{retriever}, so in the interest of keeping the number of combinations manageable, we focus on the retriever's parameters.

\subsection{Results}

Figure~\ref{fig:plaid-params} presents the results of our parameter study. The figure breaks down the effect of each parameter when balancing retrieval latency (ms/q) and either MS MARCO Dev RR@10, Dev RBO, or DL19 nDCG@1k. Each evaluation covers a different possible goal of PLAID: finding a single relevant passage, mimicking an exhaustive search, and ranking all known relevant documents. To help visually isolate the effect of each parameter, lines connect the points that keep the other two parameters constant.

From examining the figure, it is clear that \texttt{ndocs} consistently has the most substantial effect on both effectiveness and efficiency. Selecting too few documents to score (\texttt{ndocs}=256) consistently reduces effectiveness while only saving minimal latency (around 10ms/q compared to \texttt{ndocs}=1024). Meanwhile, increasing \texttt{ndocs} further to 4096 does not benefit the quality of the top 10 results (RR@10). However, the change plays a consistent and important role in improving the quality of the results further down in the ranking (RBO and nDCG@1k). Finally, increasing \texttt{ndocs}=8192 provides no additional benefits regarding search result quality or the faithfulness of the approximation to an exhaustive search, while increasing latency substantially. Based on these observations, we recommend setting \texttt{ndocs} $\in [1024,4098]$, since the benefits of the values outside this range are minimal.

The next most influential parameter is \texttt{nprobe}. As expected, increasing the number of clusters matched for each token consistently increases the latency since more candidate documents are produced and processed throughout the pipeline. Setting the value too low (\texttt{nprobe}=1 and sometimes \texttt{nprobe}=2) can often substantially reduce effectiveness, however, since documents filtered out at this stage will have no chance to be retrieved. This is especially apparent in Dev RR@10. Meanwhile, setting this value too high can reduce efficiency without yielding any gains in effectiveness.

Finally, $t_{cs}$ has the smallest effect on retrieval effectiveness, with changes to this parameter typically only adjusting the retrieval latency. This can be see by the roughly horizontal lines in Figure~\ref{fig:plaid-params}. However, as this threshold gets too high, it can have variable effects on both effectiveness and efficiency. For instance, with Dev RR@10, setting $t_{cs}=0.6$ sometimes reduces effectiveness and increases latency. Therefore, we recommend using $t_{cs}\in[0.4,0.5]$ --- and preferably towards the higher end of the range to limit the effect on latency.

We now consider the effect of all three parameters together. Achieving the Pareto frontier for PLAID involves tuning all three parameters in concert. For instance, the lowest retrieval latency requires a very low value of \texttt{ndocs}. However, lowering \texttt{ndocs} to 256 from 1024 without corresponding changes to the other parameters could simply yield worse effectiveness without making much of a dent in latency. Meanwhile, boosting \texttt{ndocs} without also adjusting \texttt{nprobe} will increase latency without improving effectiveness. Figure~\ref{fig:plaid-pareto} (on Page 1) perhaps shows the effect of this patchwork of parameters most clearly, with the Pareto frontier formed of various combinations of $\texttt{nprobe}\in\{1,2,4,8\}$, $t_{cs}\in\{0.3,0.45,0.5,0.6\}$, and $\texttt{ndocs}\in\{256,1024,4096,8192\}$.

In summary, each of PLAID's parameters plays a role in the final efficiency-effectiveness trade-offs of the algorithm. While \texttt{ndocs} plays the most important role, properly setting \texttt{nprobe} (and to a lesser extent, $t_{cs}$) is also necessary to achieve a good balance. In some ways, the importance of \texttt{ndocs} is unsurprising since the more documents you score precisely, the higher effectiveness one can expect (up to a point). But this begs some important questions. What is the impact of the source of the pool of documents for exact scoring? Is PLAID's progressive filtering process worth the computational cost compared to simpler and faster candidate generation processes? We answer these questions by exploring re-ranking baselines in the following section.

\section{Baseline Study}

The original paper compared PLAID's efficiency to three baselines: (1) Vanilla ColBERT(v2), which uses IVF indexes for each token for retrieval, in line with the method used by original ColBERT(v1)~\cite{DBLP:conf/sigir/KhattabZ20}; (2) SPLADEv2~\cite{DBLP:journals/corr/abs-2109-10086}, which is a learned sparse retriever~\cite{DBLP:conf/ecir/NguyenMY23}; and (3) BM25~\cite{DBLP:conf/trec/RobertsonWJHG94}, which is a traditional lexical retrieval model. Among these baselines, only Vanilla ColBERT(v2) represents an alternative retrieval engine; SPLADEv2 use other scoring mechanisms and act as points of reference. Curiously, the evaluation omitted the common approach of just re-ranking the results from an efficient-but-imprecise model like BM25. In this section, we compare PLAID with this baseline. Further, we compare both approaches with Lexically Accelerated Dense Retrieval (LADR)~\cite{DBLP:conf/sigir/KulkarniMGF23}, which modifies the re-ranking algorithm to also consider the nearest neighbors of the top-scoring results encountered when re-ranking.

\begin{figure*}
\centering
\includegraphics[scale=0.76]{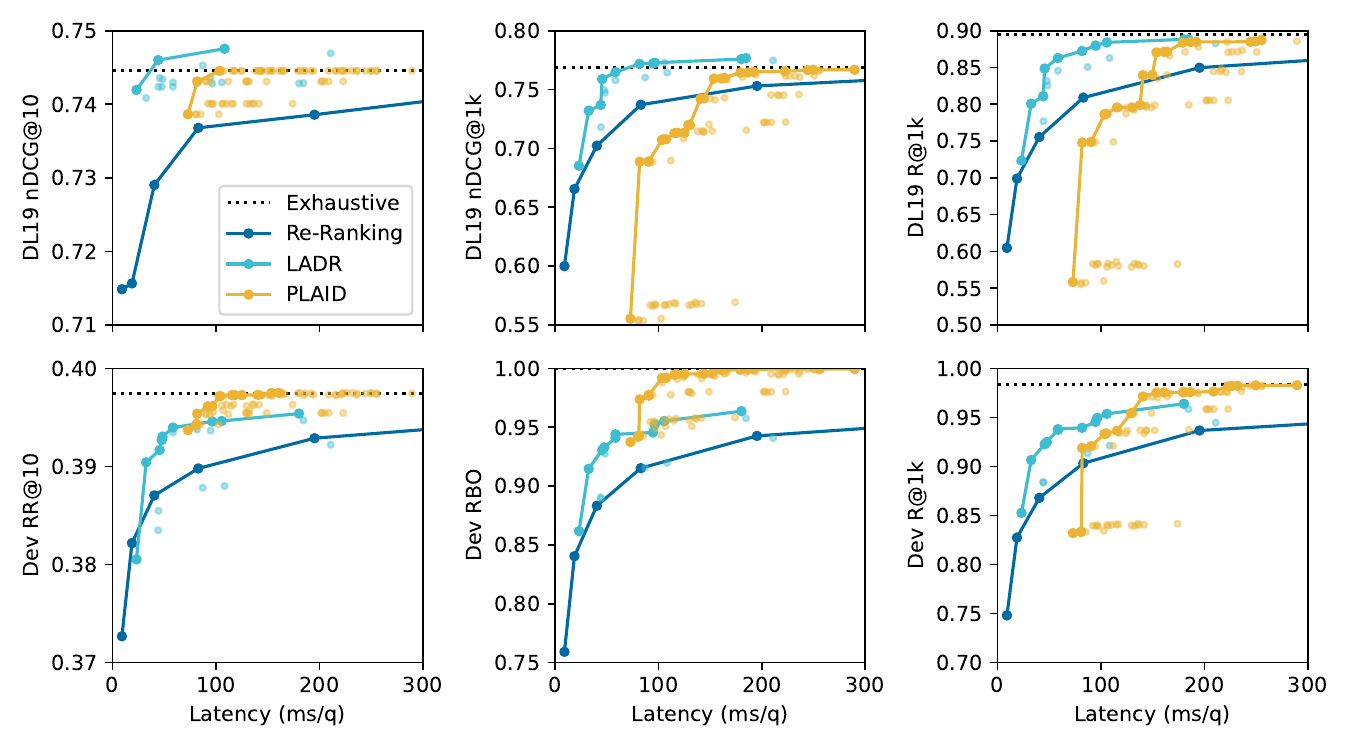}
\vspace{-2em}
\caption{Results of our baseline study. The lines connecting points for each approach represent its Pareto frontier.}
\label{fig:baselines}
\end{figure*}

\subsection{Experimental Setup}

We use PLAID's experimental results from Section~\ref{sec:params} as a starting point for our baseline study. We further modify the PLAID source code to support two more approaches: re-ranking and LADR.

\textbf{Re-Ranking.} We use the efficient PISA engine~\cite{DBLP:conf/sigir/MalliaSMS19} for BM25 retrieval, using default parameters and a BlockMaxWAND~\cite{DBLP:conf/sigir/DingS11} index structure. We then re-rank those results using ColBERTv2's decompression and scoring function. Given that we found the number of candidate documents for scoring to be the most important parameter for PLAID, we vary the number of retrieved results from BM25 as each of the following values: $n\in\{200, 500, 1000, 2000, 5000, 10000\}$. Note that due to the dynamic index pruning applied, performing initial retrieval is considerably faster for low values of $n$ than for higher ones---in addition to the cost of ColBERTv2 decompression and scoring.

\textbf{LADR.} We further build upon the re-ranker pipeline using LADR. This approach incorporates a nearest neighbor lookup for top-scoring ColBERTv2 results to overcome possible lexical mismatch from the first stage retrieval. In line with the procedure for PLAID, we perform a grid search over the number of initial BM25 candidates $n\in\{100, 500, 1000\}$ and the number of nearest neighbors to lookup $k\in\{64, 128\}$. We use the precomputed nearest neighbor graph based on BM25 from the original LADR paper. By using the \textit{adaptive} variant of LADR, we iteratively score the neighbors of the top $c\in\{10,20,50\}$ results until they converge.

\textbf{Evaluation.} We use the same datasets and evaluation measures as in Section~\ref{sec:repro_exp}. In line with this setting, we include the single-threaded first-stage retrieval latency from PISA for both additional baselines. In a multi-threaded or GPU environment, we note that this first-stage retrieval could be done in parallel with the ColBERTv2 query encoding process, further reducing the cost of these baselines. However, given the single-threaded nature of our evaluation, we treat this as additional latency.

\subsection{Results}

Figure~\ref{fig:baselines} presents the results from our baseline study. We begin by focusing on the BM25 re-ranking pipeline. We observe that this pipeline can retrieve substantially faster than the fastest PLAID pipeline (as low as 9ms/q at $n=200$, compared to 73ms/q for the fastest PLAID pipeline). Although this setting typically reduces the quality of results compared to the fastest PLAID pipelines (Dev RR@10, RBO, R@1k, and DL19 nDCG@10), it is still remarkably strong in terms of absolute effectiveness. For instance, its Dev RR@10 is 0.373 which is stronger than early BERT-based cross-encoders~\cite{DBLP:journals/corr/abs-1901-04085} and more recent learned sparse retrievers~\cite{DBLP:journals/corr/abs-2109-10086}. 

As the BM25 re-ranking pipeline considers more documents, the effectiveness gradually improves. In most cases, however, it continues to under-perform PLAID. For instance, when considering the top-10 documents via DL19 nDCG@10 and Dev RR@10, the Pareto frontier of the re-ranking pipeline always under-performs that of PLAID. Nevertheless, the low up-front dcost of performing lexical retrieval methods makes re-ranking an appealing choice when latency or computational cost are critical.

Re-ranking is inherently limited by the recall of the first stage, however, and when the first stage only enables lexical matches, this can substantially limit the potential downstream effectiveness. We observe that LADR, as an efficient pseudo-relevance feedback to a re-ranking pipeline, can largely overcome this limitation. On DL19, LADR's Pareto frontier completely eclipses PLAID's, both in terms of nDCG and recall. (LADR's non-optimal operational points are also consistently competitive.) Meanwhile, on Dev, LADR provides competitive---albeit not always optimal---effectiveness. Given that Dev has sparse assessments and DL19 has dense ones, we know that LADR is selecting suitable relevant documents as candidates, even though they are not necessarily the ones ColBERTv2 would have identified through an exhaustive search. The RBO results on Dev further reinforce this: while PLAID can achieve a nearly perfect RBO compared with an exhaustive search, LADR maxes out at around 0.96.

In summary, re-ranking and its variant LADR are highly competitive baselines compared to PLAID, especially at the low-latency settings that PLAID targets. Although they do not necessarily identify the same documents that an exhaustive ColBERTv2 search would provide, the baselines typically provide alternative documents of high relevance.

We note that re-ranking comes with downsides, however. It requires building and maintaining a lexical index alongside ColBERT's index, which adds storage costs, indexing time, and overall complexity to the retrieval system. Nonetheless, these costs are comparatively low compared to those of deploying a ColBERTv2 system itself. For instance, a ColBERTv2 index of MS MARCO v1 consumes around 22GB of storage, while a lexical PISA index uses less than 1GB. Meanwhile, hybrid retrieval systems (i.e., those that combine signals from both a lexical and a neural model) will need to incur these costs anyway. LADR adds additional costs in building and maintaining a document proximity graph (around 2GB for a graph with 64 neighbors per document on MS MARCO).

\begin{figure}
\centering
\includegraphics[scale=0.7]{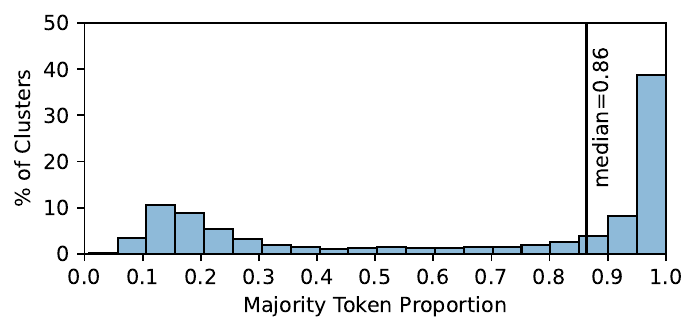}
\vspace{-1em}
\caption{The distribution of Majority Token Proportions among clusters for ColBERTv2.}
\label{fig:cluster-analysis-mtp}
\end{figure}

\subsection{Cluster Analysis}

Curious as to why re-ranking a lexical system is competitive compared to PLAID, we conduct an analysis of the token representation clusters PLAID uses for retrieval vis-à-vis the lexical form of the token. We use the ColBERTv2 MS MARCO v1 passage index from the previous experiments, and modify the source to log the original token ID alongside the cluster ID and residuals of each token. We then conduct our analysis using this mapping between the token IDs and cluster IDs.

We start by investigating how homogeneous token clusters are. In other words, we ask the question: \textit{Do most of a cluster's representations come from the same source token?} We first observe that most clusters map to multiple tokens (the median number of tokens a cluster maps to is 15, while only 2.2\% of tokens only map to a single token). However, this does not tell the complete story since the distribution of tokens within each cluster is highly skewed. To overcome this, we measure the proportion of each cluster that belongs to the majority (or plurality) token. Figure~\ref{fig:cluster-analysis-mtp} presents the distribution of the majority token proportions across all clusters. We observe that 39\% of clusters have a majority proportion above 0.95 (i.e., over 95\% of representations in these clusters come from the same token). Meanwhile, the median proportion among all clusters is 0.86. Only 2.7\% of clusters have a majority proportion less than 10\%. Collectively, these results suggest that although clusters are frequently formed of multiple tokens, they are usually heavily dominated by a single token. In other words, they largely perform lexical matching.

\begin{figure}
\centering
\includegraphics[scale=0.44]{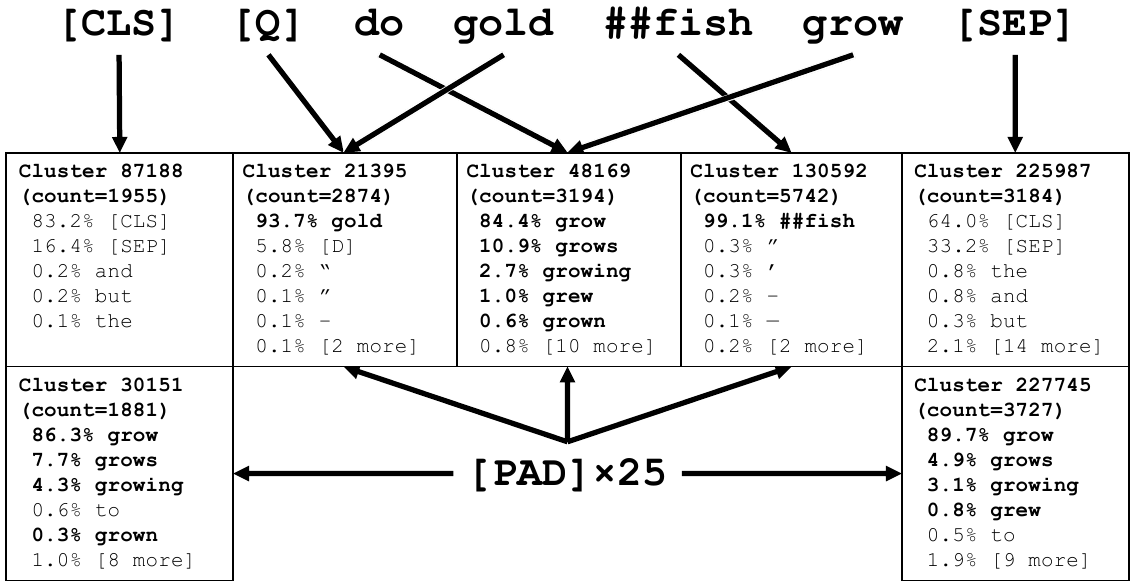}
\caption{Example query, its corresponding clusters retrieved by PLAID, and the original tokens that contributed to each cluster. Bold tokens are (stemmed) lexical matches from the query.}
\label{fig:cluster-analysis-example}
\end{figure}

Within a cluster, what exactly are the other matching tokens? Figure~\ref{fig:cluster-analysis-example} provides example clusters for the MS MARCO query ``do goldfish grow''. Some of the matching clusters (48169 and 225987) perform rather opaque semantic matching over \texttt{[CLS]} and \texttt{[SEP]} tokens. These clusters match either other such control tokens or (much less frequently) function words like \texttt{and}, \texttt{but}, and \texttt{the}. We suspect these function words are coopted to help emphasize the central points of a passage, given that they typically do not provide much in terms of semantics on their own. Next, three clusters (48169, 30151, and 227745) each have majority token proportions below or near the median. However, many of the minority tokens within a cluster are just other morphological forms of the same word: \texttt{grow}, \texttt{grows}, \texttt{growing}, etc. In other words, they share a common stem. When merging stems, these three clusters all have majority token proportions above 95\%. The final two clusters (21395 and 130592) are dominated (>90\% majority token proportion) by a single token. Like the control tokens, these pick up on punctuation tokens, which we suspect are coopted to help emphasize particularly salient tokens within a passage.
This qualitative analysis suggests that although some clusters likely perform semantic matching, Figure~\ref{fig:cluster-analysis-mtp} may actually be underestimate the overall prevalence of lexical matching among PLAID clusters.

\begin{figure}
\centering
\includegraphics[scale=0.7]{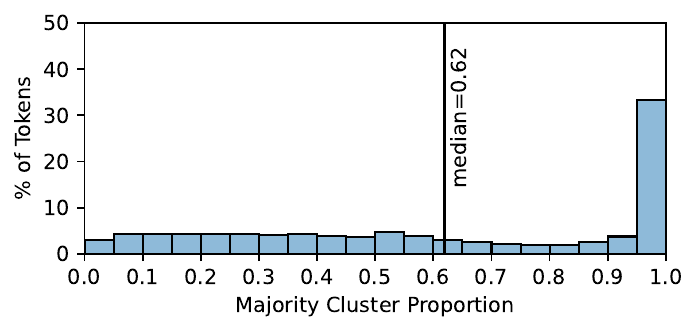}
\vspace{-1em}
\caption{The distribution of Majority Cluster Proportions among tokens for ColBERTv2.}
\label{fig:cluster-analysis-mcp}
\end{figure}

The observation that most clusters map to a single source token only tells half the story, however. Perhaps PLAID is effectively performing a form of dynamic pruning~\cite{DBLP:conf/cikm/BroderCHSZ03}, wherein query terms only match to a certain subset of lexical matches (i.e., the most semantically related ones) rather than all of them. After all, Figure~\ref{fig:cluster-analysis-example} showed three separate clusters with the same majority token (\texttt{grow}). Therefore, we ask the inverse of our first question: \textit{Do most of a token's representations map to the same cluster?} Akin to the cluster analysis, we measure the majority cluster proportion for each token, and plot the distribution in Figure~\ref{fig:cluster-analysis-mcp}. Here, 33\% of tokens have a majority cluster proportion greater than 0.95. Unlike our observations in Figure~\ref{fig:cluster-analysis-mtp}, the tail is flatter and more uniform, giving a median majority cluster proportion of 0.62. These results suggest that although a sizable number of tokens map to a single prominent cluster, many tokens are spread among many different clusters. However, as can be seen in Figure~\ref{fig:cluster-analysis-example}, just because a token appears in many different clusters doesn't mean that it will necessarily be pruned off completely: two clusters that feature \texttt{grow} (30151 and 227745) are captured directly by the \texttt{[PAD]} ``expansion'' tokens of the query.

This analysis demonstrates that PLAID performs a considerable amount of lexical matching (though not exclusively so) when identifying documents to score. It also provides some insights into why re-ranking is competitive against PLAID.

\section{Conclusion}

In this paper, we conducted a reproducibility study of PLAID, an efficient retrieval engine for ColBERTv2. We were able to reproduce the study's main results, and showed that they successfully generalize to a dataset with more complete relevance assessments. We also showed that PLAID provides an excellent approximation of an exhaustive ColBERTv2 search. Using an in-depth investigation of PLAID's parameters, we found that they are highly interdependent, and the suggested settings are not necessarily optimal. Specifically, it is almost always worth increasing \texttt{ndocs} beyond the recommended 256, given the low contribution to latency and high boost in effectiveness that the change provides. Meanwhile, the missing baseline of simply re-ranking a lexical system using ColBERTv2 (and its recent variant, LADR) provides better trade-offs in terms of efficiency and effectiveness at low-latency operational points. However, these baselines do not provide as strong of a true approximation of an exhaustive ColBERTv2 search. Finally, an analysis showed that PLAID relies heavily on lexical matches for the initial retrieval of documents.

Our study provides important operational recommendations for those looking to deploy a ColBERTv2 system, both with and without the PLAID engine. It also further highlights the importance of comparing against versatile re-ranking systems when evaluating the efficiency of retrieval algorithms. Given the indirect way that PLAID performs first-stage lexical matching, future work could investigate methods for hybrid PLAID-lexical retrieval. By relying on PLAID for semantic matches and a traditional inverted index for lexical matches, we may be able to achieve the ``best of both worlds'': the high-quality ColBERTv2 approximation of PLAID and the high efficiency of re-ranking.

\section*{Acknowledgments}

This work is supported, in part, by the Spoke ``FutureHPC \& BigData'' of the ICSC – Centro Nazionale di Ricerca in High-Performance Computing, Big Data and Quantum Computing, the Spoke ``Human-centered AI'' of the M4C2 - Investimento 1.3, Partenariato Esteso PE00000013 - "FAIR - Future Artificial Intelligence Research", funded by European Union – NextGenerationEU, the FoReLab project (Departments of Excellence), and the NEREO PRIN project funded by the Italian Ministry of Education and Research Grant no. 2022AEFHAZ.

\bibliographystyle{ACM-Reference-Format}
\bibliography{biblio}

\end{document}